\documentstyle[12pt]{article}
\title{The rare decays $B \to X_{s,d}\; \nu \bar \nu$ and $B_{s,d} 
\to l^+l^- $ in the Multiscale Walking Technicolor Model 
\thanks{Supported by the 
 National Natural Science Foundation of China under Grant No.19575015
 and By the Sino-British Friendship Scholarship Scheme.}}
\author{ Xiao Zhenjun$^{1,2}$, L\"u Linxia$^1$, Guo Hongkai$^1$, 
Lu Gongru$^1$ \\  
{\small $^1$ Department of Physics, Henan Normal University,
Xinxiang, 453002} \\
{\small $^2$ Department of Physics, Peking University, Beijing, 100871} }
\date{\today}

\begin{document}
\maketitle
\begin{abstract}
We calculate the contributions to the rare B-decays,  $B \to X_{s,d} \nu 
\bar \nu$, $B_{s,d} \to l^+l^- $ from the unit-charged technipions. 
Within the considered parameter space we find that: (a) the enhancements
to the branching ratios in question can be as large as three orders of
magnitude; (b) the ALEPH data of $B \to X_s \nu \bar \nu$ lead to strong
mass bounds on $m_{p1}$ and $m_{p8}$: 
$m_{p8} \geq 620, 475 GeV$ for $F_Q=40GeV$ and $m_{p1}=100, 400 GeV$
respectively.  (c) the CDF data of $B_s \to \mu \bar \mu$ lead to
a relatively weak limit: $m_{p8} \geq 320 GeV$ for $F_Q=40GeV$ and
$m_{p1}=200 GeV$.
\end{abstract}  

\vspace{0.5cm}

\newcommand{\beq}{\begin{eqnarray}}
\newcommand{\eeq}{\end{eqnarray}}
\newcommand{\paa}{\pi_1}
\newcommand{\pbb}{\pi_8}
\newcommand{\pap}{\pi_1^+}
\newcommand{\pam}{\pi_1^-}
\newcommand{\pbp}{\pi_8^+}
\newcommand{\pbm}{\pi_8^-}
\newcommand{\bsz}{b\overline{s}Z}
\newcommand{\bsd}{B \to X_{s,d} \; \nu \overline{\nu}}
\newcommand{\bs}{B \to X_{s} \; \nu \overline{\nu}}
\newcommand{\bd}{B \to X_{d}\; \nu \overline{\nu}}
\newcommand{\bbsd}{B_{s,d} \to l^+l^-}
\newcommand{\bbs}{B_{s} \to \mu^+  \mu^-}
\newcommand{\bbd}{B_{d} \to \mu^+ \mu^-}

\newcommand{\ssa}{\sin^2\theta_W}
\newcommand{\cca}{\cos^2\theta_W}
\newcommand{\mpa}{m_{p1}}
\newcommand{\mpb}{m_{p8}}

\newcommand{\xzxt}{X_0(x_t)}
\newcommand{\cztc}{C_0^{New}}
\newcommand{\brbs}{B(B \to  X_{s}\nu \bar \nu)}
\newcommand{\brbd}{B(B \to X_{d} \nu \bar \nu)}
\newcommand{\brbsd}{B(B \to X_{s,d}\nu\bar\nu)}
\newcommand{\brbbs}{B(B_{s} \to \mu^+  \mu^-)}
\newcommand{\brbbd}{B(B_{d} \to \mu^+  \mu^-)}

\vspace{1cm}

In the framework of the Standard Model (SM), the rare decays $\bsd$ 
and $\bbsd$ are theoretically very clean and dominated by similar 
$Z^ 0$-penguin 
and W-box diagrams involving top quark exchanges \cite{buras974,buras961}. 
These rare B-decay modes therefore may play an important role in 
searching  for the new physics beyond the SM \cite{lu96}.
    
In ref.\cite{lane91}, Lane and Ramana constructed a specific multiscale 
walking technicolor model(MWTCM) and investigated its phenomenology. 
This  model predicted a rich spectrum of technipions. Among them are 
unit-charged color-octets $\pi^a_{\bar D U}$ ( $a$ is the 
color index) and unit-charged 
color-singlets $P_1^+$ and $P_2^+$. In this letter, we use the symbol 
$\pbb$ and $\mpb$ to denote the color-octet $\pi_{\bar D U}$ and its mass. 
The mixed state $\paa$ of the 
$P_1^+$ and $P_2^+$ is  the same kind of technipion as $P^+$ given in 
ref.\cite{farhi}.  We will study the new contributions to 
the rare B-decays from the physical mixed state $\paa$ instead of the two 
technipions $P_1^+$ and $P_2^+$, for the sake of simplicity.  According
to the estimations in ref.\cite{lane91}, $\mpa \sim 200 GeV$
and $\mpb \sim 300 GeV$. 

In ref.\cite{zpc971}, the authors calculated the contributions to the
rare decay $b \to s \gamma $ due to the effective $bs\gamma$ coupling
induced by the $\paa$ and $\pbb$ appeared in the MWTCM and found that
resultant  enhancement to the branching ratio $B(B \to X_s \gamma)$ 
can be  two
orders of magnitude. The CLEO data\cite{cleo95} led to strong limit
on charged technipion masses: $\mpb > 600 GeV$ for $\mpa =300 GeV$.
The MWTCM itself is therefore
strongly disfavored by the CLEO data\cite{cleo95}. 

In this letter, we calculate the new contributions to the rare decays
$\bsd$ and $\bbsd$ due to the effective $\bsz$ coupling induced also by  
the $\paa$ and $\pbb$. This work is complementary to the relevant studies
about the new effects to the $b \to s \gamma$ decay\cite{zpc971} and the
rare  K-decays $etc$ \cite{epj981,lu96}, in order  to test or  constrain
the Technicolor(TC) models by currently available data of rare B- and 
K-decays.

In the numerical calculation, we treat $\mpa$ and $\mpb$  as
semi-free parameters, varying in the ranges of
$100 GeV \leq \mpa \leq 400 GeV$ and $200 GeV \leq \mpb \leq 800 GeV$ 
respectively. The relevant Yukawa and gauge couplings of   
charged technipions to fermion pairs and to $Z^0$ gauge boson can be 
found in refs.\cite{lu96,farhi,ellis81}. 

The new penguin diagrams for the induced $\bsz$ couplings due to 
the exchange of the $\paa$ and $\pbb$ are shown in Fig.1. 
The corresponding one-loop diagrams in the SM were evaluated long time 
ago and can be found in ref.\cite{inami81}. Only the color-singlet $\paa$
couples to lepton pairs, and therefore may contribute to the 
rare B-decays in question through the box diagrams. But
the Yukawa couplings between $\paa$ and $l \nu$ pairs are strongly
suppressed by the lightness of the lepton masses $m_l$ ($l = e \mu, \tau$).
Consequently, we can neglect  the tiny contributions from $\paa$ through 
the box diagrams safely. 

Because of the lightness of the $s$ and $b$ quarks  when 
compared with the top quark mass $m_t$ and the technipion masses 
$\mpa$ and $\mpb$ we set $m_s=0$ and $m_b=0$
in the calculation. We will use dimensional regularization to regulate 
all the ultraviolet divergences in the virtual loop corrections
and adopt the modified minimal subtracted ($\overline{MS}$) 
renormalization scheme.  

By analytical evaluations of the Feynman diagrams as shown in Fig.1, 
we find the effective $\bsz$ vertex induced by the $\paa$ and $\pbb$
exchanges,
\beq
\Gamma^I_{Z_{\mu}} = 
\frac{1}{16 \pi^2}\frac{g^3}{\cos\theta_W}\; \sum_{j} V_{js}^*V_{jb}\,
\overline{s_L}\, \gamma_{\mu}\, b_L\; C_0^{New}(y_j)
\label{bsza}    \\
\Gamma^{II}_{Z_{\mu}} = 
\frac{1}{16 \pi^2}\frac{g^3}{\cos\theta_W}\; \sum_{j} V_{js}^*V_{jb}\,
\overline{s_L}\, \gamma_{\mu}\, b_L\; C_0^{New}(z_j)
\label{bszb}
\eeq
with   
\beq
C_0^{New}(y_j) =\eta_{TC}^a \; \left[ \frac{y_j(-1 +2\ssa -3y_j 
+2\ssa y_j)}{8(1-y_j)} - \frac{\cos^2\theta_W y_j^2}{2(1-y_j)^2}\ln(y_j) 
\right]\label{cza}    \\
C_0^{New}(z_j)=\eta_{TC}^b \,\left[ 
\frac{z_j(-1 +2\ssa -3z_j +2\ssa z_j)}{8(1-z_j)} -
 \frac{\cos^2\theta_W z_j^2}{2(1-z_j)^2}\ln(z_j) \right]
\label{czb} 
\eeq
and 
\beq
\eta_{TC}^a =\frac{\mpa^2}{3\sqrt{2}F_Q^2 G_F M_W^2}, \ \  
\eta_{TC}^b =\frac{8\mpb^2}{3\sqrt{2}F_Q^2 G_F M_W^2} 
\eeq
where $y_j=m_j^2/\mpa^2$, $z_j=m_j^2/\mpb^2$, $V_{ij}$ ($i=u, c, t$ and $j=d, 
s, b$) are the  elements of the CKM mixing matrix, 
$M_W$ is the mass of W gauge
 boson, $F_Q$ is the technipion decay constant in the MWTCM, $\sin \theta_W$ 
is the Weinberg angle, and $G_F=1.16639 \times 10^{-5} (GeV^{-2})$ is the 
Fermi coupling constant.
The  functions $C_0^{New}(y_j)$ and $C_0^{New}(z_j)$ in eqs.(\ref{cza},
\ref{czb}) are  just the same kind of functions as  
the basic function
$C_0(x_i)$ in eq.(2.18) of ref.\cite{buras974}. Functions $C_0(y_t)$ and
$C_0(z_t)$ 
describe the contributions to the $\bsz$ vertex from the $\paa$ and 
$\pbb$ respectively. The $C_0(y_t)$ is always positive, but the
$C_0(z_t)$ will change its sign from $"+"$ to $"-"$ at $\mpb =531GeV$.

Within the SM, the rare B-decays under consideration depend on 
the functions $X(x_t)$ and/or $Y(x_t)$, they are currently known at the 
next-to-leading order level \cite{buras974,buras961}.
When the new contributions from charged technipions are included, one has 
\beq
X(x_t, y_t, z_t) &=& X(x_t) + C_0^{New}(y_t) + C_0^{New}(z_t),
\label{xxtt} \\
Y(x_t, y_t, z_t) &=& Y(x_t) + C_0^{New}(y_t) + C_0^{New}(z_t). 
\label{yxtt}
\eeq
where $x_t=m_t^2/m_W^2$, $y_t=m_t^2/\mpa^2$ and $z_t=m_t^2/\mpb^2$. 

In the following numerical calculations, we  fix the relevant parameters 
as follows and use them as the Standard Input: $ 
M_W=80.2GeV$, $\alpha_{em}=1/129$, $\ssa =0.23$, $
m_t \equiv \overline{m_t}(m_t) = 170GeV$, $\tau(B_s)=\tau(B_{d})=1.6ps$, 
$\Lambda^{(5)}_{\overline{MS}}=0.225GeV$, $F_{B_s}=0.210GeV$, 
$m_{B_s}=5.38GeV$, $ m_{B_d}=5.28GeV$, $A=0.84$, $\lambda=0.22$, 
$\rho=0$ and $ \eta=0.36$. For the definitions and values of these input 
parameters, one can see refs.\cite{buras974,buras961}.

Within the SM, normalizing to the semi-leptonic branching ratio 
$B(B \to X_{c} e \overline{\nu})$ and summing over the three 
neutrino flavors one finds\cite{buras974}
\beq
B(\bs) = B(B \to X_{c} e \overline{\nu}) \frac{3 \alpha_{em}^2}
{4 \pi^2\sin^4\theta_{W}} \frac{\mid V_{ts} \mid ^2}{\mid V_{cb} \mid ^2} 
\frac{X(x_{t})^2}{f(z)} \frac{\overline{\eta}}{\kappa(z)} 
\label{brs1}
\eeq
where $\bar{\eta}=\kappa(0)$, $f(z)$ and $\kappa(z)$ with $z=m_c/m_b =0.29$ 
are the phase-space and quantum chromodynamics(QCD) correction factors for 
the decay $B \to X_{c} e \overline{\nu}$,
\beq
f(z)&=& 1- 8z^2 + 8 z^6 - z^8 -24 z^4 \ln(z),\\
\kappa(z)&=& 1- \frac{2\alpha_s(m_b)}{3\pi} \left[ (\pi^2 - \frac{31}{4})
(1-z)^2 + \frac{3}{2} \right]
\eeq
where $\alpha_s(m_b)$ is the QCD coupling constant at the energy scale 
$\mu=m_b$.

Within the SM, using the Standard input parameters and setting 
$B(B \to X_{c}e \overline{\nu})=10.4\%$ and $\mid V_{ts}/V_{cb} \mid ^2=0.95$,
one finds $\brbs=3.52 \times 10^{-5}$ and $\brbd=2.03 \times 10^{-6}$.
When the new contributions due to $\paa$ and $\pbb$ are included, 
one has $\brbs=1.57\times 10^{-3}$ for the typical values of $F_Q=40GeV$,
$m_{p1}=200 GeV$ and $m_{p8}=300GeV$, that is two-orders of magnitude 
higher than the value of the SM prediction.  The color-octet $\pbb$ dominates the total contribution.

In Fig.2  the dot-dash (solid, dots) curve shows the branching ratio
when the new contributions from $\paa$ and $\pbb$ are taken into
account for $F_Q=40 GeV$ and  $m_{p1}=100GeV$ ($200 GeV$, $400GeV$).
The horizontal short-dash line shows the 
ALEPH bound on the branching ratio $\brbs$ \cite{aleph96}: $\brbs < 7.7
\times 10^{-4}$, which is a factor of 20 above the SM expectation but
sensitive enough to put stringent limits on technipion masses. Assuming 
$F_Q=40GeV$, one has $m_{p8} > 620 GeV$ for $m_{p1}=100GeV$,
$530GeV < \mpb < 890 GeV$ for $\mpa =200GeV$ and $475 GeV < \mpb < 740 GeV$
for $\mpa =400GeV$.  For smaller $F_Q$, the constraints will become much
stronger.

In the case of the decay $\bd$ one has to replace $V_{ts}$ in eq.(\ref{brs1})
by $V_{td}$,  which results in a decrease of the branching ratio by roughly 
an order of magnitude. But unfortunately, no experimental bound on
the decay $\bd$ is available currently.

Within the SM, using the effective Hamiltonian as given in 
ref.\cite{buras974} one finds
\begin{eqnarray}
B(B_s \to \l^+ l^-)=
\frac{\tau(B_s)G_F^2}{\pi} (\frac{\alpha_{em}}{4 \pi \ssa})^2
F_{B_s}^2 m^2_{l} m_{B_s} \sqrt{1-\frac{4 m_{l}^2}{m^2_{B_{s}}}}
\mid V_{tb}^{*} V_{ts} \mid^2 Y(x_t)^2
\end{eqnarray}
with s replaced by d in the case of $B_d \to l^+ l^-$.

\begin{table}
\caption{The branching ratios of $\bbsd$ for l=e,$\mu$,$\tau$.}
\begin{center}
\begin{tabular}{|c|c|c|c|}
\hline
{\rm Branching ratio} & {SM} & {SM + New}& {Data} \\
\hline
{$B(B_{s} \to e^+ e^-) $}      & $0.73\times 10^{-13}$ &{$1.9\times
10^{-10} - 1.7 \times 10^{-12}$}& \\ \hline
{$B(B_{s} \to \mu^+ \mu^-)$}   & $3.13\times 10^{-9}$ & {$8.1\times
10^{-6} - 7.3 \times 10^{-8}$}& {$\leq 2.0 \times 10^{-6}$} \\ \hline
{$B(B_{s} \to \tau^+ \tau^-)$} & $0.67\times 10^{-6}$ & {$1.7\times
10^{-3} - 1.5 \times 10^{-5}$} & \\ \hline \hline
{$B(B_{d} \to e^+ e^-) $}      & $0.44\times 10^{-14}$ & {$1.2\times
10^{-12} - 1.0 \times 10^{-14}$}& \\ \hline
{$B(B_{d} \to \mu^+ \mu^-)$}   & $1.90\times 10^{-10}$ & {$4.9\times
10^{-8} - 4.4 \times 10^{-10}$}&{$\leq 6.8 \times 10^{-7}$} \\ \hline
{$B(B_{d} \to \tau^+ \tau^-)$} & $0.40\times 10^{-7}$  & {$1.1\times
10^{-5} - 0.9 \times 10^{-7}$}& \\ \hline
\end{tabular}
\end{center}
\end{table}

In the numerical calculations, we use the standard input parameters and
assume that $F_Q=40GeV$, $\mpa = 200GeV$, $200GeV \leq \mpb \leq 800GeV$, 
and set $\mid V^{*}_{tb}V_{ts} \mid ^2=0.0021$ and $\mid V^{*}_{tb}V_{td} 
\mid ^2=1.3 \times 10^{-4}$.  The numerical results with the inclusion of
new contributions from the technipions $\paa$ and $\pbb$ are listed in
Table 1.

For the decays $B_s \to l^+ l^-$, the available experimental bound is
$\brbbs \leq 2.0 \times 10^{-6}$\cite{cdf98}, which leads to the
lower bounds on $\mpb$: $\mpb > 320 GeV$ for $F_Q=40 GeV$ and $\mpa=200
GeV$. But this bound is much weaker than that from the ALEPH data of
$B \to X_s \nu \bar \nu$. 
For the decays $B_d \to l^+ l^-$, the available experimental bound is
$\brbbd \leq 6.8 \times 10^{-7}$ \cite{cdf98},  which is  
still not sensitive enough to put any limits on $\mpa$ and $\mpb$.

In summary, the ALEPH data of $B\to X_s \nu \bar \nu $ lead to strong limits
on the charged technipion masses $\mpa$ and $\mpb$. The  assumed mass ranges
of $\paa$ and  $\pbb$ in the MWTCM \cite{lane91} are excluded 
and therefore the model itself is strongly disfavored by ALEPH
data\cite{aleph96}. Other relevant
studies\cite{zpc971,epj981} also led to the similar conclusions. The major
problem of the MWTCM is that the heavy top quark mass is assumed to be
generated by extended technicolor interaction, which  is clearly 
unreasonable\cite{lane96}.


\vspace{1cm}
\begin{center}
{\bf Figure Captions}
\end{center}
\begin{description}

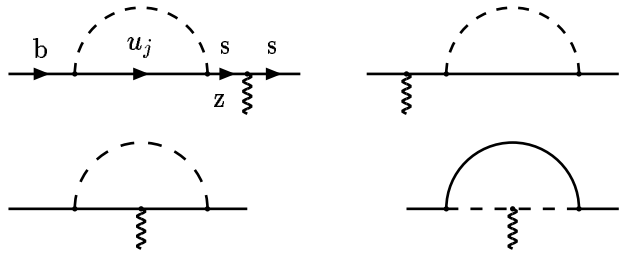
\item[Fig.1:] New $Z^0-$penguin diagrams contributing to the induced 
$\bsz$ vertex from the internal exchanges of the technipion 
$\paa$ and $\pbb$. The dashed lines are $\paa$ and $\pbb$ lines and  
$u_j$ stands for the quarks $(u, c, t)$.

\item[Fig.2:] The $m_{p8}$ dependence of 
the branching ratio $\brbs$ when the new contributions are included.
The horizontal short-dash line shows the ALEPH upper bound, while   
the dot-dash (solid, dots) curve shows the branching ratio for 
$m_{p1}=100GeV$ ($200 GeV$, $400GeV$).

\end{description}

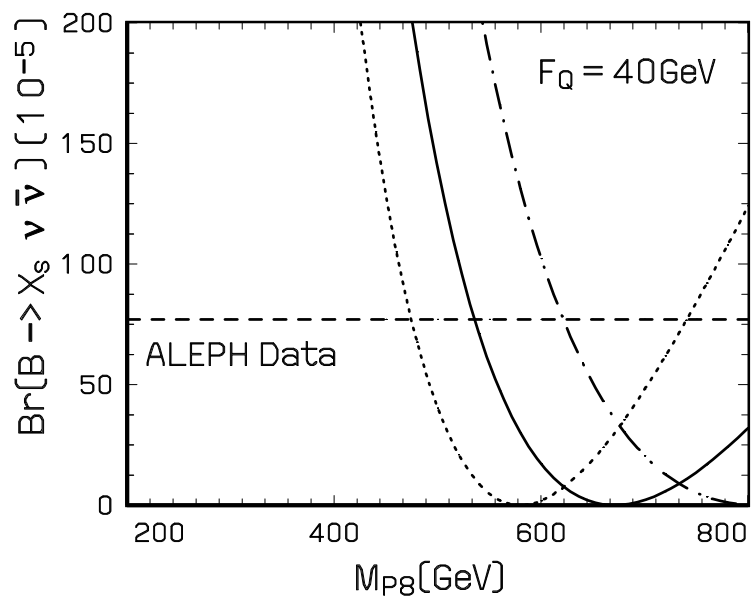
\end{document}